\documentstyle[preprint,aps]{revtex}
\begin{document}
\draft
\title{\bf {How general is Legett's conjecture for a mesoscopic
ring? }}
\author{P. Singha Deo\cite{eml} }
\address{Institute of Physics, Bhubaneswar 751005, India}
\maketitle
\begin{abstract}
It has been shown[19] that in a loop of length u to which a
single stub of length v is attached (fig. 1 in ref. 19), the
parity effect is completely destroyed when v/u$>2$. It was also
shown that such minute topological defects (v/u$<<1$) act as
singular perturbations. However ref. 19 studies the effect of a
single topological defect and says that for v/u$<<1$ parity
effect is not violated in the ring. In this paper we show that
topological defects of the type v/u$<<1$ can also violate the
parity effect depending on the exact value of v/u and the parity
effect is significantly destroyed if we have many such geometric
scatterers. This paper brings out the physical reasons for the
destruction of parity effect. We show that the generic feature
of topological defects as this is that they can produce
discontinuous phase change (with change in
energy) of the electron wavefunction in the
ring for special value of v/u and then Legett's cojecture breaks
down. So Legett's conjecture which generalises parity effect in
presence of any arbitrary 2 body scattering and any arbitrary 1
body scattering need not be true when the one body scattering
can produce discontinuous phase change of the electron wave
function in the ring. This may have implications on the
experimental observations.
\end{abstract}
\narrowtext
\newpage

A normal metal ring pierced by a magnetic field carries a
persistent current and has a magnetic response. The ring shows
strong parity effect in the sense that the nature of response of
a ring (paramagnetic or diamagnetic) is exteremely sensitive to
the number of electrons present in the ring. For spinless
electrons, a clean ring with odd number of electrons has a
diamagnetic response and that with an even number of electrons
has paramagnetic response[1]. This happens because when the
number of electrons in the ring changes from odd to even, there
is a statistical half flux quantum  which shifts the energy flux
dependence by exactly $\phi_{0}$/2[2]. It was conjectured by
Legett that this fact follows just from symmetry property of the
wavefunction (the electrons being fermions, the wave function
must be antisymmetric) and is independent of electron electron
interaction and impurity scattering[3]. Refs. [4,5,6] are
devoted to proving rigorously the so called Legett's conjecture.
At high temperature and disorder the parity effect shows up as a
shift by $\pi$ of the persistent current versus flux curve with
the change of number of electrons from odd to even[1,2]. For
electrons with spin we get double parity effect[2]. Only in case
of electrons with spin as well as interactions the parity effect
may be destroyed because of fractional Aharonov Bohm(AB)
effect[7-12].  But such fractional Aharonov Bohm effect has not
yet been observed experimentally. Parity effect is seen in
multichannel simulations too[13].

The first experiment[14] was done with N=$10^7$ rings and due to
parity effect one expects the ensemble averaged response to
scale with $\sqrt N$.  Contradictory to it the actual response
measured in the experiment is quiet high along with a $\phi_0/2$
periodicity.  There are many possible effcts that can give the
$\phi_0/2$ periodicity[15] but the magnitude is a puzzle[16] and
so it is for a single ring (where alternate levels have opposite
response) experiment[17]. In a recent single loop experiment[18]
one observes fair agreement.

It has been shown[19] that in a loop of length u to which a
single stub of length v is attached (fig. 1 in ref. 19), the
parity effect is completely destroyed when v/u$>2$. It was also
shown that minute topological defects (v/u$<<1$) act as singular
perturbations. However ref. 19 studies the effect of a single
topological defect and says that for v/u$<<1$ parity effect is
not violated in the ring. In this paper we show that topological
defects of the type v/u$<<1$ can also violate the parity effect
depending on the exact value of v/u and the parity effect is
significantly destroyed if we have many such geometric
scatterers and that can well affect the experimental
observations. This paper brings out the physical reasons for the
destruction of parity effect. We show that the generic feature
of topological defects as this is that they can produce
discontinuous phase change of the electron wavefunction in the
ring for special value of v/u and then Legett's cojecture breaks
down. So Legett's conjecture which generalises parity effect in
presence of any arbitrary 2 body scattering and any arbitrary 1
body scattering need not be true when the one body scattering
can produce discontinuous phase change of the electron wave
function in the ring.

The thickness of the experimental ring could not have been
uniform. If there are some sharp variations in thickness then
some resonant cavities may be formed at certain places (at
random) in the ring. 1-D modelling of the ring helps to
understand the basic physics and its result can be easily
extended to the multichannel case. Resonant cavities can be
taken as stubs[20] and width of the resonant cavitites result
only in lowering the energy[21].

The allowed modes in the system are given by the following
condition[19].

\begin{equation}
cos(\alpha)={sin(ku)cot(kv)\over 2} + cos(ku)
\end{equation}

\noindent k is the allowed wave vectors and
$\alpha= 2\pi\phi/\phi_{0}$, is the AB phase, $\phi$ being the
flux through the ring and $\phi_{0}$ is the flux quantum. The
above condition is the simplified form of the following
condition[19].

\begin{equation}
cos(\alpha)=re(1/T)
\end{equation}

\noindent where T is the transmission amplitude across the ring
when the ring is cut open. The above equation immediately
suggests something. For the clean ring the bound state condition
is $e^{i(ku+\alpha)}=1$. Whereas eqn(2) is just the condition
(It is worth mentioning that $cos(\alpha)=cos(2n\pi-\alpha)$)

\begin{equation}
e^{i[cos^{-1}(re{1\over T})+\alpha]}=1
\end{equation}

B$\ddot u$ttiker et al[22] has shown that an electron in a ring
with a random potential is effectively moving in a periodic
system whose unit cell is the ring when cut open.  It is also
well known that $cos^{-1}(re{1\over T})$ is the Bloch phase (Ku
where K is the Block momentum) aquired by the electron in
traversing an unit cell of an infinite periodic system where T
is the transmission amplitude across the unit cell of the
periodic system[19].  Hence eqn(3) is just $e^{i(Ku+\alpha)}$=1
and is in perfect agreement with ref. [22].  It also suggests
that inside the ring the electron moves with the momentum K and
not with the free particle momentum $\pm k$. Hence it is not
surprising that $\pm k$ states are degenerate even in the
presence of magnetic field (because both +k and -k satisfy eqn
(3) ). Inside the ring the electron is moving anticlockwise or
clockwise with momentum K (Ku=$cos^{-1}(re{1 \over T})$).  One
of them is a diamagnetic (anticlockwise moving) state and the
other is a paramagnetic (clockwise moving) state.  Initially as
the magnetic field is increased then the two states move away
from each other, however the diamagnetic state and the
paramagnetic state are not degenarate for any value of $\phi$
for reasons explained later.  In a clean ring (ku+$\alpha$) is
the phase aquired by the electron in going round the ring once
whereas in a ring with scatterers (potential or geometric)
$Ku+\alpha$ is the phase aquired in moving round the ring once
and eqn (3) is due to the single valuedness of the wavefunction.
So eqn (2) suggests that a particular mode is allowed in the
ring if the phase aquired by an electron wave fn. in that mode
(apart from the AB phase) in going round the ring once,i.e.
$cos^{-1}(re{1 \over T})$, equals ($\alpha$). We can alternately
state it as - if the Block phase of an electron in travelling a
unit cell of an infinite periodic system i.e., Ku equals
$\alpha$ then single valuedness of wavefunction is obtained in
the ring made of the unit cell and we get a bound state.  So the
boundstates can be determined by graphically solving
re(1/T)=cos($\alpha$).  This occurs at certain k values and then
$k^2$ is the energy of the electron in that particular state K.
It has an analogy with a scattering problem where k is the
momentum outside the potential where it vanishes, whereas the
momentum K inside the potential can be quiet different. However
the energy throughout is $k^2$ (we have set $\hbar=1$ and 2m=1).

So in fig(1) we show a simple plot where the solid curve is a
plot of y=re(1/T) with ku for v/u=.2. Wherever this curve
intersects the straight line y=cos($\alpha$), the corresponding
k value is a bound state for the system. Let us start with
$\alpha$=0 and then y=cos(0) curve is shown in fig. 1 by dotted
lines. Two consequitive points where the curve y=re(1/T)
intersects the straight line y=cos(0) are denoted by A and B in
the fig. The corresponding k values are denoted as $k_1$ and
$k_2$ in the fig. If $\alpha$ is increased gradually then the
straight curve y=cos($\alpha$) shifts gradually downwards
towards the dashed curve. As the curve y=cos($\alpha$) gradually
go downwards the allowed wave vectors $k_1$ and $k_2$ slowly
drift rightwards and leftwards respectively along the k axis. As
$k_1$ drifts rightwards with $\alpha$ i.e., towards higher
energy, $k_1$ is a diamagnetic state. Similarly $k_2$ is a
paramagnetic state.  That $k_1$, $k_2$, etc. gradually increase
or decrease with $\alpha$ gives rise to a dispersion with
$\alpha$ (E vs $\alpha$) with close by alternate states going
further away from each other with $\alpha$ upto $\alpha= \pi$.
y=cos($\pi$) is also shown in fig. 1 with dashed lines. If we
increase $\alpha$ further then the straight curve
y=cos($\alpha$) start moving upwards and comes back to its
original position at $\alpha =2 \pi$. This ensures $\phi_0$
periodicity of the dispersion curves. Since cos$(\alpha$) can
vary from -1 to +1 (dotted lines to dashed lines) the dispersion
curve for any two consequtive states can never cross (see fig.
1). So the dispersion curve is exactly similar to that of a ring
with a random potential.  The cause of gaps in the dispersion
curve in that case is the breakdown of rotational symmetry of
the ring by the random potential and hence the removal of
degenaracy of states that cross over for a clean ring. In our
case the rotational symmetry is destroyed by the topological
defect. Some gaps (the ones around kv=$n \pi$) are very large
but most gaps are very small. In fact some special gaps may
actually go to zero for reasons explained later.  Hence from
fig. 1 it is evidient  that alternate states carry persistent
currents with opposite signs and have opposite magnetic
properties up to infinite energy. This is exactly the same as in
case of potential scattering.

But this effect is not observed when we plot the same curves for
different values of v/u=.21 (fig. 2) (in fact v/u=.2$\pm
\epsilon$ is sufficient). Consider the intersections between the
graphs y=re(1/T) and y=cos(0).  The first few alternate states
have opposite magnetic properties but the fifth and the sixth
states (two consequtive states marked A and B) are both
diamagnetic disobeying the parity effect. Slowly increase
$\alpha$ to see that. Parity effect is again violated for the
11th and the 12th states both of which are paramagnetic. After a
regular spacing of five levels we always find two consequtive
levels that violate the parity effect.  Hence parity effect due
to the antisymmetric property of the electron wave function is
not generic to a ring with topological defects. We shall soon
see why it does not happen for specific values of v/u.

Note that for kv=n$\pi$, the transmission across a stub is
zero[23] due to the formation of a node at the junction between
the ring and the stub. This mode always lie in a gap of the
dispersion curve and is never an allowed mode.

Scattering by a topological defect like a stub is still a poorly
understood phenomenon. Ref. [23] tries to explain scattering by
a stub on the same footing as scattering by potentials. Here we
intend to understand the problem by mapping it onto an effective
delta potential. A special feature of the delta potential is
that $\mid T \mid ^2$= re(T). This feature is not seen for any
other potential. However this feature is also observed in case
of a stub.  This makes it possible to map a single stub onto an
effective delta potential V(x)=k cot(kv)$\delta$(x).  So the
strength of the delta potentials depend on the fermi energy.
That is why at certain energies V(x) becomes zero and then the
gap vanishes. Now let us start with k=0 and then slowly increase
k continuously. For k=0 V(x)=1/v which means it starts with a
small positive value. Then it decreases and soon goes to zero.
After this the strength of the potential monotonously increase
on the negative side and finally becomes -$\infty$ at kv=$\pi$.
After this V(x) undergo a discontinuous jump from $-\infty$ to
$+\infty$. If the strength of the $\delta$ potential at kv=$\pi$
and kv=$\pi + \epsilon$ are discontinuous the scattering phase
shift and hence the Block phase will also undergo discontinuous
jump. re(1/T) also make a discontinuous jump from $-\infty$ to
$\infty$ and hence the Block phase jumps by $\pi$ (see fig. 2)
(Block phase of the infinite periodic system has to be defined
to a modulo of $2
\pi$ i.e., $-1<re(1/T)<1$).
The next allowed Block phase of the infinite periodic system of
stubs after that at D is that at B and they differ by $\pi$.
This is markedly different from the next allowed Block phase at
any other gap, e.g., the Block phase at C is same as that at A.
We have seen that if the Block phase of the periodic system of
stubs equals the AB phase $\alpha$ (for the time being we have
taken $\alpha$=0) then the single valuedness of the wave
function gets satisfied in the ring and we get a bound state.
This additional phase results in satisfying this condition and
creating a state at B close to the value kv=$\pi $ which
otherwise would not have been there had the phase change across
kv=$\pi$ been continuous. If it so happens that this singularity
in the Bloch phase due to a singularity in the effective
potential V(x) is cancelled by another singularity then the
phase difference between two consequitive Block phase would not
have been $\pi$ and then this state at B would not exist because
total phase aquired in this state is not enough to satisfy the
single valuedness condition in the ring. All other states would
have been as usual and would have been qualitatively same as
that of a ring with a random potential. This is what happens in
case of fig. 1. For v/u=.2 at kv=$n \pi$ cot(kv)=$\pm \infty$
but sin(ku)=0. And so there is no discontinuty in re(1/T).
Hence the state at B of fig. 2 will not exist. See fig. 3 where
we have superposed the two graphs of fig. 1 and fig. 2. The
dotted curve is for v/u=.2 whereas the solid curve is for
v/u=.21. All the states for both the cases are very close to
each other except that the solid curve shows a state at A (which
is the state at B in fig. 2) where the dashed curve does not
show any state at all. Other states of the solid curve are very
close but slightly at lower energies than those of the dashed
curve because an increase in v/u means an increase in the phase
space of the electrons. The state at A has no partner and locked
between a diamagnetic state and a paramagnetic state it has to
break the parity effect. It is easy to see from fig 2 and 3 that
slope of re(1/T) is such that if it jumps from $-\infty$ to
$+\infty$ then the broken parity state is diamagnetic and
paramagnetic for the other case.  Specific values of the
parameter v/u at which these two singularities exactly cancell
are negligibly few compared to the values where they do not and
is hardly likely in a real situation.

We then study the spectrum of a ring with four small topological
defects or stubs present in it. To find the spectrum we have to
solve eqn (2) numerically using the transfer matrix mechanism to
compute T[24]. A portion of plot is shown in figs. 4a, 4b (solid
lines). Within a certain energy range (3400$>E u^2 >$150) there
are much more diamagnetic levels than paramagnetic. Higher above
there are however more paramagnetic levels than diamagnetic.
The length of the stubs as well as the separation between the
stubs has been chosen by random number genarating subroutines
(.1$<v/u<$.3). We have plotted for only one configuration
because we do not intend to take an average over configurations
and compare with the experimentally observed numbers, because
our calculations are in 1-D where localisation effects are very
strong.  But the usual alternate paramagnetic and diamagnetic
states apart from the parity breaking states are almost
unaltered by the defects because they feel a very weak V(x). We
have checked that the graphs qualitatively remain the same for
other configurations and for each case there is a substantial
breakdown of parity effect.  In a 3D multichannel ring there are
many subbands and very close by levels. Each subband exhibits
the parity effect[3]. Topological defects will destroy the
parity effect of each subband with the first one being
diamagnetic for each.  Even one appropriate topological defect
in that case can give many broken parity states.

The purpose of this paper has been to show the breakdown of
Legett's conjecture and parity effect due to topological defects
(v/u$<<$1). We have also shown that a discontinuity in the
scattering phase is the physical reason for it because nature of
a state depend on how its boundary conditions get tuned by the
phases. The discontinuty in the phase shifts the origin of the
total phase by $\pi$ but does not determine wheather the total
phase should increase or decrease with $\alpha$ and k.  However
we would like to say that any breakdown in parity effect is sure
to enhance the persistent current in the single ring as well as
in the many ring experiment specially when the usual states (not
the parity breaking ones) are not affected much by the
topological defects.  Also initially upto a certain energy there
is no breakdown of parity effect which means that the persistent
current in a ring with very few propagating channels (only four
in ref. 18) will hardly be affected by these defects. Also the
abundant paramagnetic states much higher up may not be occupied
and hence may be of no consequence. No theories proposed so far
try to find a common explanation for the three experiments. Also
all except one[GK in 16] completely overlook the fact that the
rings can exhibit coexistence of large persistent currents and
small conductances[17].  In our case as for every parity
breaking state there is a state at which the conductance across
the ring becomes zero ($\mid T \mid ^2$=0) the same reason that
enhances the persistent current can decrease the conductance.

v/u values taken in this paper for qualitative analysis are much
higher than realistic values.  Smaller is v/u the higher up will
be the diamagnetic states.  Again more the width of the resonant
cavities lower again will be the parity breaking states[22]. A
3D simulation is needed to study the real situation and it will
be reported in the near future.

The author thanks Prof. A. M. Jayannavar for usefull
discussions. The author
also acknowledges Prof. A. M. Srivastava for a brief discussion
on topology.

\vfill
\eject

\vfil
\eject

\centerline {\bf FIGURE CAPTIONS}

Fig. 1. To show graphical solutions for the allowed modes for
v/u=.2.

Fig. 2. To show the graphical solutions for the allowed modes
for v/u=.21.

Fig. 3. Superposition of fig. 1. and fig. 2.

Fig. 4a. E versus $\phi$ dispersion curves in the range
$1000<Eu^2<2000$.

Fig. 4b. E versus $\phi$ dispersion curves in the range
$2000<Eu^2<3400$.

\vfill
\eject
\end{document}